\documentclass[prd,onecolumn,notitlepage,nofootinbib,superscriptaddress,showpacs,showkeys]{revtex4-2}

\usepackage{amsfonts}
\usepackage{graphicx}
\usepackage{dcolumn}
\usepackage{amsmath}
\usepackage{amssymb}
\usepackage[T1]{fontenc}
\usepackage[utf8]{inputenc}
\usepackage{esint}
\usepackage[brazil,english]{babel}
\usepackage[usenames,dvipsnames]{color}
\usepackage{longtable}
\usepackage{ulem}
\usepackage{nicefrac}
\usepackage{bm}
\usepackage{tikz}
\usepackage{braket}
\usepackage{graphicx}

\begin{document}

\title{Black-Hole Thermodynamics from Gauge Freedom in Extended Iyer--Wald Formalism}

\author{Thiago de L. Campos}
\email{thiagocampos@usp.br}
\affiliation{Instituto de F\'{\i}sica, Universidade de S\~{a}o Paulo, Caixa Postal 66318,
05315-970, S\~{a}o Paulo-SP, Brazil}

\author{Mario C. Baldiotti}
\email{baldiotti@uel.br}
\affiliation{Departamento de F\'{\i}sica, Universidade Estadual de Londrina, CEP
86051-990, Londrina-PR, Brazil}

\author{C. Molina}
\email{cmolina@usp.br}
\affiliation{Escola de Artes, Ci\^{e}ncias e Humanidades, Universidade de S\~{a}o Paulo, Avenida Arlindo Bettio 1000, CEP 03828-000, S\~{a}o Paulo-SP, Brazil}

\begin{abstract}
Thermodynamic systems admit multiple equivalent descriptions related by transformations that preserve their fundamental structure. This work focuses on exact isohomogeneous transformations (EITs), a class of mappings that keep fixed the set of independent variables of the thermodynamic potential, while preserving both the original homogeneity and the validity of a first law. Our investigation explores EITs within the extended Iyer--Wald formalism for theories containing free parameters (e.g., the cosmological constant). EITs provide a unifying framework for reconciling the diverse formulations of Kerr-anti de Sitter (KadS) thermodynamics found in the literature. While the Iyer--Wald formalism is a powerful tool for deriving first laws for black holes, it typically yields a non-integrable mass variation that prevents its identification as a proper thermodynamic potential. To address this issue, we investigate an extended Iyer--Wald formalism where mass and thermodynamic volume become gauge dependent. Within this framework, we identify the gauge choices and Killing vector normalizations that are compatible with EITs, ensuring consistent first laws. As a key application, we demonstrate how conventional KadS thermodynamics emerges as a special case of our generalized approach.
\end{abstract}

\maketitle

\section{Introduction}

It is well established that a physical system can be described by multiple thermodynamic potentials, each corresponding to a particular choice of independent variables. These equivalent descriptions are related through specific maps, such as Legendre transformations, which exchange roles of extensive and intensive quantities, while preserving thermodynamic consistency. 
A deeper geometric perspective arises from contact geometry~\cite{bln}, where the thermodynamic phase space is formulated as a contact manifold. In this framework, the equivalence between different thermodynamic potentials is naturally encoded through contactomorphisms, transformations that preserve the underlying contact structure. This approach not only generalizes classical Legendre duality but also provides a unified mathematical foundation for thermodynamic covariance.

In black-hole thermodynamics, the Iyer--Wald formalism~\cite{iw,wald1993black} provides a powerful framework for deriving first laws in arbitrary diffeomorphism-invariant gravitational theories. However, this approach typically yields a variation of the mass $(\delta M)$ that is non-integrable~\cite{hs}, preventing its identification as the differential of a proper \mbox{thermodynamic potential. }

Recent work on Kerr-anti de Sitter (KadS) spacetime showed that choosing a Killing field that considers the angular velocity of the black hole with respect to infinity results in a non-integrable mass variation~\cite{xiao2024extended}. However, an exact differential can still be extracted from this selection.
The procedure introduces an additional term that explains the difference between thermodynamic and geometric volumes~\cite{xiao2024extended}. 
But there is a drawback. This method relies, to some extent, on prior knowledge of the expected first law, as the resulting $\delta M$ should (ideally) correspond to the variation of a proper thermodynamic potential from \mbox{the outset.}

In the present work, we investigate an extension of the Iyer--Wald formalism for gravitational theories with free parameters, such as the cosmological constant. In our framework, both the mass and the thermodynamic volume of the black hole are gauge dependent. 
More precisely, these quantities are defined through the potential volume, a concept commonly employed in KadS thermodynamics~\cite{kastor2009enthalpy}.  
The fact that this volume is gauge dependent was noted~\cite{kmt} but its consequences were not fully exploited.
This gauge dependence supplements the usual freedom in the normalization of the Killing generator of the event horizon.

Additionally, we show that thermodynamic potentials obtained
via Legendre transformations compose a particular subset of a broader class of transformations, which we have defined as exact isohomogeneous transformations (EITs)~\cite{nos}. These mappings act on thermodynamic representations preserving (i) the set of independent variables of the original potential, (ii) homogeneity, and (iii) the validity of the first law.

We have demonstrated~\cite{nos} that a notable application of EITs is in providing a unifying framework to understand the different thermodynamic descriptions of KadS black holes found in the literature~\cite{gpp,hawking1999rotation,gao2024general}. In the current study, our goal is to systematically identify the gauge transformations, and the corresponding transformations of the Killing vector, that are compatible to the EITs, thereby ensuring consistent first laws of thermodynamics through the extended Iyer--Wald formalism.

This paper is organized as follows. In Section~\ref{sec:gen}, the general formalism for generating different thermodynamic descriptions is presented. Isohomogeneous transformations are introduced, with special emphasis on the so-called exact ones and how they are formalized in the context of contact geometry. In Section~\ref{sec:gauge free}, the gauge-dependent extension of the Iyer--Wald formalism is discussed. This framework is combined with EITs to generate integrable first laws in Section~\ref{sec: gauge transf}, and it is applied to Kerr-anti de Sitter black holes in Section~\ref{sec: KadS}. Final comments are presented in Section~\ref{sec: finals}. Additionally, Appendix~\ref{app:contact} provides an overview of contact manifolds, Appendix~\ref{app: non} briefly reviews nonexact isohomogeneous transformations, and Appendix~\ref{app:pot} expands a discussion on potential and vector volumes.

\section{Generating Thermodynamic Descriptions}

\label{sec:gen}

\subsection{Isohomogeneous Transformations}

\label{sec it}

The thermodynamic description of a system is specified by an exact 1-form, whose integration is usually called a thermodynamic potential. The symmetry of this representation is encoded in the homogeneity of this potential. 
Formal setups where thermodynamic theories can be established are the symplectic~\cite{bfm} and the contact~\cite{bln} geometries. These frameworks allow the connection between thermodynamics and other physical theories~\cite{brav}. 

As the dynamics generated by a symplectic manifold are invariant under canonical transformations (symplectomorphisms), submanifolds describing the equilibrium states of a thermodynamic system are invariant under the so-called contactomorphisms (i.e., contact symmetries).
Well-known examples of contactomorphisms are the Legendre transformations, which relate different thermodynamic potentials (internal energy, enthalpy, etc.). A broader type of contactomorphism is given by the exact isohomogeneous transformations, which we introduced in~\cite{nos}.

Given a thermodynamic potential $M_{0}$ satisfying a first law and an Euler relation, 
\begin{equation}
\begin{split}
	&\mathrm{d}M_{0} =T_{0}\,\mathrm{d}S+\sum_{i=2}Y_{0}^{i}\,\mathrm{d}X_{i} ~, 
	\\
	&rM_{0} =\alpha _{1}ST_{0}+\alpha _{2}X_{2}Y_{0}^{2}+\alpha
	_{3}X_{3}Y_{0}^{3}+\cdots ~,
\end{split}
\end{equation}
we define an isohomogeneous transformation via a pair of functions $(g,h)$ as follows:
\begin{equation}
	M_{1}\equiv gM_{0}~,\quad T_{1}\equiv gT_{0}+h\frac{\partial g}{\partial S}~,\quad
	Y_{1}^{i}\equiv gY_{0}^{i}+h\frac{\partial g}{\partial X_{i}}~.  \label{tig}
\end{equation}
In Equation~\eqref{tig}, $h$ is a homogeneous function of the same degree $r$ as $M_{0}$, and $g $ is a homogeneous function of degree zero: 
\begin{equation}
\begin{split}
	& rh=\alpha _{1}S\frac{\partial h}{\partial S}+\alpha _{2}X_{2}\frac{	\partial h}{\partial X_{2}}+\alpha _{3}X_{3}\frac{\partial h}{\partial X_{3}}+\cdots ~, 
	\\
	& \alpha _{1}S\frac{\partial g}{\partial S}+\alpha _{2}X_{2}\frac{\partial g}{\partial X_{2}}+\alpha _{3}X_{3}\frac{\partial g}{\partial X_{3}}+\cdots
	=0 ~.
\end{split}
\end{equation}

The new function $M_{1}$ (which is also homogeneous of degree $r$) satisfies the relations
\begin{equation}
\begin{split}
	&  T_{1}\,\mathrm{d}S+\sum_{i=2}Y_{1}^{i}\,\mathrm{d}X_{i}\,=\mathrm{d}	M_{1}+\left( h-M_{0}\right) \mathrm{d}g ~,
	\\
	& rM_{1}=\alpha _{1}ST_{1}+\alpha _{2}X_{2}Y_{1}^{2}+\alpha
	_{3}X_{3}Y_{1}^{3}+\cdots ~.
\end{split}
\label{int}
\end{equation}
From Equation~\eqref{int}, it is apparent that $M_{1}$ is an exact differential (i.e., defines a new thermodynamic potential) if and only if $g$ is constant or $h=M_{0}$. The latter defines what we term exact isohomogeneous transformations. We emphasize that, despite the homogeneity of $M_{1}$, for a nonexact homogeneous transformation, the second relation in Equation~\eqref{int} does not represent an Euler relation. In black-hole thermodynamics, Equation~\eqref{int} can usually be identified with a Smarr formula, derived from generalized Komar integrals. 

The link between dimensional analysis and homogeneity often supports the interpretation of the Smarr relation as the geometric counterpart of the Euler relation. This type of scaling argument is employed in~\cite{kastor2009enthalpy} to justify treating the cosmological constant as a pressure term. However, this identification should be used with caution.
While a purely geometric approach for black-hole thermodynamics may lead to the second relation in Equation~\eqref{int} in some particular cases, the isohomogeneous transformations ensure that this is an Euler relation for $M_1$ only when it is obtained from a proper thermodynamic description via Equation~\eqref{tig}. Otherwise, the set $(M_1,T_1, Y_1^i)$ does not, in general, provide a valid thermodynamic representation. An important example is the Smarr formula for KadS which employs Hawking’s results~\cite{hawking1999rotation} and does not lead to a first law~\cite{gpp}. Additional comments on this topic can be found in Appendix~\ref{app: non}.
Therefore, the distinction between the Smarr formula and the Euler relation is a central point in the construction of an adequate thermodynamic description~\cite{nos}.

In what follows, we examine EITs in more detail, emphasizing their
connection to contactomorphisms on contact manifolds, and show how several well-known constructions from standard thermodynamics arise as special cases of such transformations. In Appendix~\ref{app: non}, we also explore
nonexact transformations, which can be employed to restore the integrability
of a first law in certain scenarios~\cite{nos}.

\subsection{Exact Isohomogeneous Transformations as Contactomorphisms}

Exact isohomogeneous transformations allow the construction of equivalent thermodynamic descriptions. In the context of black-hole thermodynamics, 
these transformations enable the connection of various formulations found in the literature~\cite{nos}. However, the formal structure of these transformations, as well as their relation to those used in conventional thermodynamics, have not yet been fully explored. This is the goal of the \mbox{present section.}

Appendix~\ref{app:contact} provides a concise introduction to contact
geometry, including relevant definitions and general results used in the present work. 
Given a contact manifold $\left( \mathcal{T},\eta \right) $, with contact 1-form $\eta $, thermodynamic phase space $\mathcal{T}$, and canonical coordinates $(M_{1},X_{i},Y_{1}^{i})$ for an EIT (exact isohomogeneous transformation), we have, from Equation~\eqref{tig}, with $h = M_0$, and Equation~\eqref{c1}, the following:  
\begin{equation}
	\eta =\mathrm{d}M_{1}-\sum_{i=1}Y_{1}^{i}\mathrm{d}X_{i}=g\left[ \mathrm{d} 
	M_{0}-\sum_{i=1}Y_{0}^{i}\mathrm{d}X_{i}\right] =g\alpha ~,  \label{nec}
\end{equation}%
where we identify $X_{1}=S$ and $Y_{0}^{1}=T$. Thus, as an isomorphism $\mathrm{EIT}:\mathcal{T}\rightarrow \mathcal{T}$ in
the thermodynamic phase space $\mathcal{T}$, we have: 
\begin{equation}
	(\mathrm{EIT})^{\ast }\alpha =g\alpha ~,
\end{equation}%
which shows that an EIT is a contactomorphism. Moreover, because $\alpha$
and $\eta $ share the same kernel, they yield equivalent thermodynamic
descriptions. Since $g\neq 1$, this contactomorphism is not strict.

In standard thermodynamics, where we work with intensive and extensive variables (especially with temperature, which is intensive, i.e., homogeneous of degree zero), we can choose $g=\nicefrac{1}{T_0}$. In this case,
\begin{equation}
	\eta =\frac{1}{T_{0}}\mathrm{d}M_{0}-\mathrm{d}S-\sum_{i=2}\frac{Y_{0}^{i}}{T_{0}}\mathrm{d}X_{i}~.
\end{equation}
The associated Reeb vector field \eqref{reeb} is 
\begin{equation}
	\xi =-\frac{\partial }{\partial S}~,
\end{equation}
corresponding to the entropy representation. Since $T_{0}$ is intensive, this representation is equivalent to the EIT with $h=M_{0}$ and $g=1/T_{0}$.
Nevertheless, they define different thermodynamic descriptions, which differ in the choice of variables used to describe the system.%
\footnote{As detailed in Appendix~\ref{app:contact}, formally this means that the two descriptions differ in the choice of vertical subspace $L_{\xi }$ of the distribution $D$ that decomposes the tangent bundle $T\mathcal{T}=L_{\xi	}\oplus D$.} 
In the map $\mathrm{EIT}:\mathcal{T}\rightarrow \mathcal{T}$,
the new Reeb vector is given by 
\begin{equation}
\xi =\frac{\partial }{\partial M_{1}}~.
\end{equation}
If $T_{0}$ is not intensive, the thermodynamic descriptions are not related, since
the entropy and energy representations differ in their homogeneity degree.

Certain special cases of EITs are closely related to standard Legendre transformations. These represent EITs given by
\begin{equation}
h=M_{0}~,\quad g=g_{L}\equiv 1-\sum_{k\in I}\frac{X_{k}Y_{0}^{k}}{M_{0}}~,
\end{equation}
where $I$ is an arbitrary set of indices. Under this transformation, 
\begin{equation}
\mathrm{d}M_{1}=\sum_{i=1}Y_{1}^{i}\mathrm{d}X_{i}=\sum_{i\notin I}Y_{0}^{i}%
	\mathrm{d}X_{i}-\sum_{k\in I}X_{k}\mathrm{d}Y_{0}^{k}~,
\end{equation}
with 
\begin{equation}
Y_{1}^{i}=g_{L}Y_{0}^{i}+M_{0}\frac{\partial g_{L}}{\partial X_{i}}%
=Y_{0}^{i}-\sum_{k\in I}X_{k}\frac{\partial Y_{0}^{k}}{\partial X_{i}}%
-\sum_{k\in I}Y_{0}^{k}\delta _{k}^{i}~.
\end{equation}
The description induced by $g_{L}$ admits two equivalent forms: one using
original variables $\left\{ X_{K}\right\} $ and new conjugates $\left\{
Y_{1}^{i}\right\} $, and another using the original $\{Y_{0}^{i}\}$ but with
the new variables $\{X_{k}\rightarrow Y_{0}^{k}\}$ for $k\in I$. The latter
is the traditional thermodynamic representation obtained by Legendre
transformations, while the former is the EIT representation given by $g_L$.
They describe the same thermodynamic potential ${(M_1 = g_L M_0)}$ in
different coordinates. Although equivalent, the EIT defines a thermodynamic
description that is structurally distinct from the Legendre transformation due to the different choice of independent variables. In particular,
the latter is a strict contactomorphism~\cite{bln}, while the EIT~\eqref{nec}
is not.

A special case is the total Legendre transformation, that is, when $I$ includes all independent variables. In this case,
\begin{eqnarray}
	&&Y_{1}^{i}=-\sum_{k=1}X_{k}\frac{\partial Y_{0}^{k}}{\partial X_{i}}%
	=-\sum_{k=1}X_{k}g^{ik}~,\quad g^{ik}=\frac{\partial Y_{0}^{k}}{\partial X_{i}}=%
	\frac{\partial ^{2}M_{0}}{\partial X_{i}\partial X_{k}}~,  \label{gik} \\
	&&M_{1}=M_{0}-\sum_{k=1}Y_{0}^{k}X_{k}=-\sum_{k=1}\frac{\beta _{k}}{r}%
	Y_{0}^{k}X_{k}~,\quad \beta _{k}=r-\alpha _{k}~, \\
	&&\mathrm{d}M_{1}=\sum_{k=1}Y_{1}^{k}\mathrm{d}X_{k}=-\sum_{k=1}X_{k}\mathrm{%
		d}Y_{0}^{k}~.
\end{eqnarray}
In standard thermodynamics with extensive variables ($r=\alpha _{i}=1$), we find $M_{1}=0$, and the familiar Gibbs--Duhem relation follows:
\begin{equation}
	r=\alpha _{i}=1\implies \sum_{k=1}X_{k}\mathrm{d}Y_{0}^{k}=0~.  \label{gd}
\end{equation}
However, if any of the homogeneity degrees are nonstandard (as in the case of black-hole thermodynamics~\cite{bfm2, fbfm}), $M_{1}$ does not vanish and can be
interpreted as describing the system in the dual space of the cotangent
space. For instance, in the thermodynamic description of KadS black holes
where the cosmological constant is not included in the phase space, the
function $M_{1}$ obtained via a total Legendre transformation represents the grand thermodynamic
potential, which plays a central role in the Euclidean action formalism \cite%
{gibbons1977action}.

The matrix $g^{ik}$ in Equation~\eqref{gik} is known as the Weinhold metric~\cite{Wei}. In this case, the EIT can be interpreted as a musical isomorphism (an index-raising operation). In addition to endowing the contact manifold with a metric structure, the introduction of $g^{ik}$ connects the contactomorphism to a gauge transformation~\cite{bln}.

\section{Gauge Freedom in Extended Iyer--Wald Formalism}

\label{sec:gauge free}

\subsection{General Setup}

The Iyer--Wald formalism provides a general framework for deriving the first law of black hole thermodynamics in arbitrary diffeomorphism-invariant gravitational theories~\cite{iw}. 
In this work, we explore how the gauge freedom generated by the function $g$ in an EIT~\eqref{nec} is related to the gauge freedom in the IW formalism.

Given the Lagrangian density $L$ and the volume form $\star 1$, the variation of $\mathbf{L}=L\star 1$ with respect to a set of dynamical fields $\phi $, which include the metric, is conveniently \mbox{written as }
\begin{equation}
	\delta \mathbf{L}=\mathbf{E}\left( \phi \right) \delta \phi +\mathrm{d}%
	\boldsymbol{\Theta }\left( \phi ,\delta \phi \right) ~,
\end{equation}
with equations of motion $\mathbf{E}(\phi )$, and total derivative $\mathrm{d}\boldsymbol{\Theta }(\phi ,\delta \phi )$. 
However, from a thermodynamic point of view, the first law obtained from the traditional Iyer--Wald formalism can pose two problems. 
First, there is the (already noted) possible failure of integrability. 
Moreover, for gravitational theories
containing free parameters (such as a cosmological constant), the resulting equations of state are generally non-homogeneous.
To deal with this issue, one approach is to treat the parameters in the Lagrangian $L$ as thermodynamic variables, allowing them to vary~\cite{cfn}.
In this section, we show how this parametric variation can induce changes in the Noether--Wald charges within a gauge-dependent extension of the Iyer--Wald formalism.

If the Lagrangian also depends on a set of parameters $p_{(\alpha )}$ that are allowed to vary, we can introduce a new variation $\hat{\delta}$:
\begin{equation}
	\hat{\delta}\mathbf{L}\equiv \frac{\partial \mathbf{L}}{\partial p_{\left(
			\alpha \right) }}\delta p_{\left( \alpha \right) }~.
\end{equation}
The generalized variation of the Lagrangian is then 
\begin{equation}
	\delta \mathbf{L}=\mathbf{E}\left( \phi \right) \delta \phi +\mathrm{d}%
	\boldsymbol{\Theta }\left( \phi ,\delta \phi \right) +\frac{\partial \mathbf{%
			L}}{\partial p_{\left( \alpha \right) }}\delta p_{\left( \alpha \right) }~.
\end{equation}
With the contribution of $\hat{\delta}$, the infinitesimal Noether--Wald charge,
\begin{equation}
\bar{\delta}\mathbf{Q}_{K}-\imath_{K}\boldsymbol{\Theta }\left( \phi ,\delta \phi
\right) ~,  \label{inf N-W}
\end{equation}
satisfies the extended Iyer--Wald relation 
\begin{equation}
	\mathrm{d}\left[ \bar{\delta}\mathbf{Q}_{K}-\imath_{K}\boldsymbol{\Theta }\left(
	\phi ,\delta \phi \right) \right] =-\imath_{K}\frac{\partial \mathbf{L}}{\partial
		p_{\left( \alpha \right) }}\delta p_{\left( \alpha \right) }~.  
        \label{IWG}
\end{equation}
In Equation~\eqref{IWG}, $K$ is a Killing vector, $\bar{\delta}$ denotes a variation keeping $K$ fixed, $\imath_{K}$ is the interior product, and $\mathbf{Q}_{K}$ is a Noether charge relative (on-shell) to the exact differential form
\begin{equation}
	\mathbf{J}_{K}\equiv \boldsymbol{\Theta }\left( \phi ,\mathcal{L}_{K}\phi
	\right) -\imath_{K}\mathbf{L}~,
\end{equation}%
with Lie derivative $\mathcal{L}_{K}$.

An important case occurs when there is a set of forms 
$\left\{ \boldsymbol{\omega}_{K}^{(\alpha )} \right\}$, labeled by the index $\alpha$,~satisfying
\begin{equation}
	\imath_{K}\frac{\partial \mathbf{L}}{\partial p_{(\alpha )}}=\mathrm{d}%
	\boldsymbol{\omega }_{K}^{\left( \alpha \right) }\implies \imath_{K}\frac{%
		\partial \mathbf{L}}{\partial p_{\left( \alpha \right) }}\delta p_{\left(
		\alpha \right) }=\mathrm{d}\left[ \boldsymbol{\omega }_{K}^{(\alpha )}\delta
	p_{(\alpha )}\right] ~. 
	\label{int w}
\end{equation}
In this scenario,%
\footnote{As previously mentioned, introducing parameters in the Lagrangian variations creates new challenges for the integrability conditions and imposes additional requirements on the symmetries of the symplectic structure. Although these exigences can be explicitly analyzed in various important cases (see, for example,~\cite{hs}), they cannot be generally specified for all diffeomorphism-invariant Lagrangians. Similarly, the additional restriction imposed by Equation~\eqref{int w} must be analyzed on a \mbox{case-by-case basis.}}
we can redefine the infinitesimal Noether--Wald charge~\eqref{inf N-W} as
\begin{equation}
	\bar{\delta}\boldsymbol{Q}_{K}-\imath_{K}\boldsymbol{\Theta }+\delta p_{\left(
		\alpha \right) }\boldsymbol{\omega }_{K}^{\left( \alpha \right) }~,
\end{equation}
so that it is again conserved: 
\begin{equation}
	\mathrm{d}\left[ \bar{\delta}\boldsymbol{Q}_{K}-\imath_{K}\boldsymbol{\Theta }%
	+\delta p_{(\alpha )}\boldsymbol{\omega }_{K}^{(\alpha )}\right] =0~.
\end{equation}

We have reached an important point. The differential forms $\boldsymbol{\omega }_{K}^{(\alpha )}$
introduce a gauge freedom into the formalism, as we are free to add closed forms $\boldsymbol{\lambda }^{(\alpha )}$: 
\begin{equation}
	\boldsymbol{\omega }_{K}^{\left( \alpha \right) }\longrightarrow \boldsymbol{%
		\omega }_{K}^{\left( \alpha \right) }+\boldsymbol{\lambda }^{\left( \alpha
		\right) }~.  \label{add lambda}
\end{equation}
We seek to analyze how this gauge freedom affects the thermodynamic
description of \mbox{the system.}

\subsection{Rotating and Stationary Black Holes with a Cosmological Constant}

We examine a concrete physical system consisting of a stationary, rotating black hole with a non-vanishing cosmological constant~$\Lambda$.
We assume that the unique free parameter in the theory is $\Lambda$. 
Since the black hole is rotating and stationary, there is a Killing field normal to the horizon with the form
\begin{equation}
K=\xi +\Omega\varphi ~,
\label{eq:killing}
\end{equation}
where $\xi$ and $\varphi$ are the generators of time and angular isometries, and $\Omega$ is the angular velocity of the black hole.
In this setup, Equation~\eqref{IWG} becomes 
\begin{equation}
	\mathrm{d}[\bar{\delta}\boldsymbol{Q}_{K}-\imath_{K}\boldsymbol{\Theta }(\phi
	,\delta \phi )]=\frac{\delta \Lambda }{8\pi }\imath_{K}(\star 1)~.
	\label{Iyer-Wald eq Gao}
\end{equation}

The vector volume with respect to the Killing field $K$ in Equation~\eqref{eq:killing} is defined as~\cite{bl} 
\begin{equation}
V_{K}=\int_{\Sigma }\imath_{K}\left( \star 1\right) ~.
\label{eq:vector-volume}
\end{equation}
In Equation~\eqref{eq:vector-volume}, the hypersurface $\Sigma$ is bounded by two 2-surfaces: the horizon cross-section $\sigma_H$ and a spacelike 2-surface $\sigma$. Because  $\imath_{\varphi }(\star 1)$ vanishes on $\Sigma$, the vector volume $V_{K}$ \mbox{reduces to}
\begin{equation}
	V_{K}=\int_{\Sigma }\imath_{K}\left( \star 1\right) =\int_{\Sigma }\imath_{\xi }\left(
	\star 1\right) =V_{\xi }~.
\end{equation}
However, since $\xi $ is a Killing field,
\begin{equation}
\mathrm{d}\left[ \imath_{\xi }\left( \star 1\right) \right] =\nabla _{\mu }\xi^{\mu }\star 1 = 0 ~.
\end{equation}
Therefore, locally, from Poincar\'{e}'s lemma, there is a differential form $\star \boldsymbol{\omega }$ such that
\begin{equation}
	\imath_{\xi }\left( \star 1\right) =\mathrm{d}\star \boldsymbol{\omega }~,
	\label{origin of w}
\end{equation}%
where the Hodge dual of $\boldsymbol{\omega }$ is used for later
convenience. Hence, Equation~\eqref{Iyer-Wald eq Gao} becomes 
\begin{equation}
	\mathrm{d}\left[ \bar{\delta}\boldsymbol{Q}_{K}-\imath_{K}\boldsymbol{\Theta }%
	\left( \phi ,\delta \phi \right) +\delta P\star \boldsymbol{\omega }\right]
	=0~,\quad P\equiv -\frac{\Lambda }{8\pi }~.
    \label{eq:def-P}
\end{equation}
In Equation~\eqref{eq:def-P}, we have defined the vacuum pressure $P$ in terms of the cosmological constant $\Lambda$ in the standard way.

In this extended Iyer--Wald formalism, the variation of the black-hole energy is
\begin{equation}
	\delta M\equiv \int_{\sigma }\left[ \bar{\delta}\boldsymbol{Q}_{\xi }-\imath_{\xi
	}\boldsymbol{\Theta }(\phi ,\delta \phi )+\delta P\star \boldsymbol{\omega }%
	\right] ~,  \label{Mpot}
\end{equation}
where $\sigma $ is a 2-dimensional surface surrounding the black hole 
(for example, using  Boyer--Lindquist-type coordinates, $\sigma = \{(t_0,r_0,\theta,\phi)\}$, with $t_0$ and $r_0$ constants, and $r_0$ \mbox{sufficiently large). }

For $\sigma =\sigma _{H}$, the potential volume of the black hole~\cite{kastor2009enthalpy} is given by
\begin{equation}
	V=\int_{\sigma _{H}}\star \boldsymbol{\omega } ~.
\end{equation}
Also, the variation of angular momentum does not change in this extended formalism:
\begin{equation}
	\delta J=-\int_{\sigma }\left[ \bar{\delta}\boldsymbol{Q}_\varphi-\imath_{\varphi
	}\boldsymbol{\Theta }\left( \phi ,\delta \phi \right) \right] =-\int_{\sigma
	}\bar{\delta}\boldsymbol{Q}_\varphi~.
    \label{eq:delta-J}
\end{equation}
The second equality in Equation~\eqref{eq:delta-J} follows from the vanishing pullback of $\imath_{\varphi}\boldsymbol{\Theta}$ on $\sigma$.

The first law of thermodynamics can be structured through the integral relation:
\begin{equation}
	\int_{\sigma }\left[ \bar{\delta}\boldsymbol{Q}_{K}-\imath_{K}\boldsymbol{\Theta }%
	\left( \phi ,\delta \phi \right) +\delta P\star \boldsymbol{\omega }\right]
	=\delta M-\Omega \delta J~.
    \label{eq:integral-relation}
\end{equation}
Considering the gauge $\delta \xi ^{\mu }=0=\delta \varphi ^{\mu }$~\cite{bardeen1973four}, and evaluating the left-hand side of expression~\eqref{eq:integral-relation} with $\sigma =\sigma _{H}$, we obtain
\begin{equation}
	\int_{\sigma _{H}}\left[ \bar{\delta}\boldsymbol{Q}_{K}+\delta P\star 
	\boldsymbol{\omega }\right] =\delta \left( \frac{\kappa A}{8\pi }\right)
	+J\delta \Omega +V\delta P~,\quad \int_{\sigma _{H}}\imath_{K}\boldsymbol{\Theta }%
	\left( \phi ,\delta \phi \right) =\frac{A}{8\pi }\delta \kappa +J\delta
	\Omega ~,  \label{int gauge}
\end{equation}
where $A$ is the area of $\sigma_H$ and $\kappa$ is the surface gravity relative to $K$. Stokes' theorem allows one to combine these results as
\begin{equation}
	\delta M=\frac{\kappa }{2\pi }\delta \left( \frac{A}{4}\right) +\Omega
	\delta J+V\delta P~. 
    \label{IW gener}
\end{equation}

An important point to consider is the issue of integrability mentioned earlier. In particular, the variation of the new parameters introduces an additional complication in the integrability conditions of Equation~\eqref{IW
	gener}. This can be seen, for example, in~\cite{xiao2024extended},
where the mass variation is non-integrable, and a total variation is then extracted from the non-integrable result, which is interpreted as the actual variation of the black-hole mass. The additional term from the non-integrable factor gives rise to the well-known difference between the thermodynamic and geometric volumes of KadS black holes~\cite{dolan2011pressure,dolan2012pdv,cvetivc2011black}.

\section{Gauge Transformation in Extended Black-Hole Thermodynamics}

\label{sec: gauge transf}

We will develop a systematic procedure to derive an integrable first law through gauge-dependent redefinitions of $\delta M$ and $V$, where integrability is ensured by appropriate gauge choices. Since exact isohomogeneous transformations naturally generate distinct first-law formulations, we will translate this mechanism into the geometric framework of the extended Iyer--Wald formalism. This approach eliminates the need for ad hoc modifications to achieve integrability~\cite{xiao2024extended,gao2024general}.

The gauge freedom in Equation~\eqref{add lambda} makes $\delta M$ and $V$ ambiguous until a specific gauge is chosen. As noted in~\cite{kmt, cvetivc2011black} and developed in~\cite{rodriguez2022first}, the best that can be done is to fix this gauge so that it reproduces a desired value of $M$, which itself must be determined independently. 
We will show that, through a 
systematic gauge-fixing procedure, it is possible to guarantee the integrability of $\delta M$ and, thus, to transform Equation~\eqref{IW
	gener} into a proper first law of thermodynamics.

More precisely, exact isohomogeneous transformations provide a method to derive alternative thermodynamic descriptions, starting from an initial formulation (labeled $0$) to a new representation (labeled $1$). In this section, we establish how to implement this transformation within the extended Iyer--Wald formalism through (i) a modification of the Killing field normalization, and (ii) a
gauge transformation of the potential volume.

Let the Killing field $K_{0}$ generate a proper first law within the extended Iyer--Wald formalism, as
\begin{equation}
	\mathrm{d}M_{0}=\frac{\kappa _{0}}{2\pi }\mathrm{d}\left( \frac{A}{4}\right)
	+\Omega _{0}\mathrm{d}J+V_{0}\mathrm{d}P~,  \label{ifl}
\end{equation}
where we replace $\delta $ with $\mathrm{d}$ for explicit exactness of the differential (i.e., explicit integrability). Consider the following transformation on the Killing field: 
\begin{equation}
	K_{0}=\xi _{0}+\Omega _{0}\varphi \longrightarrow aK_{0}=\xi _{1}+\Omega
	_{1}\varphi \equiv K_{1}  \label{K0 to K1}
\end{equation}
where
\begin{equation}
	\xi _{1}\equiv a\xi _{0}+b\varphi _{0}~,\quad \Omega _{1}\equiv a\Omega _{0}-b~.
    \label{eq:xi1-omega1}
\end{equation}
The quantities $a$ and $b$ are scalars in the
spacetime coordinates, to be fixed as the analysis proceeds, and $K_{1}$ is fixed with respect to $\bar{\delta}$.
The extended Iyer--Wald formalism for the transformed Killing field is 
\begin{equation}
	\int_{\sigma }\left[ \bar{\delta}\boldsymbol{Q}_{K_{1}}-\imath_{K_{1}}\boldsymbol{%
		\Theta }\left( \phi ,\delta \phi \right) +\delta P\star \boldsymbol{\omega }%
	_{1}\right] =\delta M_{1}-\Omega _{1}\delta J~,
\end{equation}
where $\delta M_{1}$ and $V_{1}$ are defined as 
\begin{equation}
	\delta M_{1}\equiv \int_{\sigma }\left[ \bar{\delta}\boldsymbol{Q}_{\xi
		_{1}}-\imath_{\xi _{1}}\boldsymbol{\Theta }(\phi ,\delta \phi )+\delta P\star 
	\boldsymbol{\omega }_{1}\right] ~,\quad V_{1}\equiv \int_{\sigma _{H}}\star 
	\boldsymbol{\omega }_{1}~.  \label{eq1}
\end{equation}

Still working in the gauge%
\footnote{It is also possible to choose the gauge $\delta \xi _{1}^{\mu }=0$, and Equation~\eqref{int gauge0} would have the form of Equation~\eqref{int gauge} with $\kappa_{1}$, $\Omega _{1}$, and $V_{1}$.}
$\delta \xi _{0}^{\mu }=0=\delta \varphi ^{\mu }$,
the following integrals are derived:
\begin{equation}
	\int_{\sigma _{H}}\left[ \bar{\delta}\boldsymbol{Q}_{K_{1}}+\delta P\star 
	\boldsymbol{\omega }_{1}\right] =a \delta \left( \frac{\kappa _{0}A}{8\pi }
	\right) + a J\delta \left( \Omega _{0}\right) +V_{1}\delta P~, 
\end{equation}
\begin{equation}
    \int_{\sigma
		_{H}}\imath_{K_{1}}\boldsymbol{\Theta }\left( \phi ,\delta \phi \right) =\frac{1}{%
		8\pi }aA\delta \kappa _{0}+aJ\delta \Omega _{0}~.
        \label{int gauge0}
\end{equation}
After straightforward simplifications using the surface gravity $\kappa_1 = a\kappa_0$ associated with the Killing field $K_1$, we obtain
\begin{equation}
	\delta M_{1}=\frac{\kappa _{1}}{2\pi }\delta \left( \frac{A}{4}\right)
	+\Omega _{1}\delta J+V_{1}\delta P~.  \label{transformed IW first law}
\end{equation}
The exact isohomogeneous transformations ${h=M}_{{0}}$ in Equation~\eqref{tig} ensure thermodynamic consistency. That is, if the original description satisfies a first law, the transformed quantities will also obey a first law under the mapping
\begin{equation}
	M_{1}=gM_{0}~,\quad T_{1}=gT_{0}+M_{0}\frac{\partial g}{\partial S}~,\quad \Omega
	_{1}=g\Omega _{0}+M_{0}\frac{\partial g}{\partial J}~,\quad V_{1}=gV_{0}+M_{0}%
	\frac{\partial g}{\partial P}~,  \label{transformed quantities}
\end{equation}
where $g=g(S,J,P)$ is a homogeneous function of degree zero.

Furthermore, requiring the temperature to be proportional to the surface gravity determines $a$ uniquely as
\begin{equation}
	a=g+\frac{M_{0}}{T_{0}}\frac{\partial g}{\partial S}~.  \label{a}
\end{equation}
Given this expression for $a$ and the transformed angular velocity $\Omega_1$ from Equation~\eqref{transformed quantities}, the parameter $b$ is consequently fixed to
\begin{equation}
	b=M_{0}\left( \frac{\Omega _{0}}{T_{0}}\frac{\partial g}{\partial S}-\frac{%
		\partial g}{\partial J}\right) ~.  \label{b}
\end{equation}

However, the determination of $a$ and $b$ alone is insufficient to fix the volume $V$.
While the rescaling $K_{0}\rightarrow aK_{0}$ induces a proportional transformation on $\boldsymbol{\omega }_{0}$, the complete determination of $\omega_1$ requires additional gauge fixing. That is, the full transformation on $\boldsymbol{\omega}_{0}$ must include a gauge transformation:
\begin{equation}
	\star \boldsymbol{\omega }_{0}\longrightarrow a\star \boldsymbol{\omega }%
	_{0}+\star \boldsymbol{\lambda }\equiv \star \boldsymbol{\omega }_{1}~.
	\label{w0 to w1}
\end{equation}
Nevertheless, we can determine $\boldsymbol{\lambda}$ such that $V_{1}$ in Equation~\eqref{transformed quantities} is obtained from the integration of $\boldsymbol{\omega }_{1}$. More precisely, 
\begin{equation}
	V_{1}=\int_{\sigma _{H}}(a\star \boldsymbol{\omega }_{0}+\star \boldsymbol{%
		\lambda })=\left( g+\frac{M_{0}}{T_{0}}\frac{\partial g}{\partial S}\right)
	V_{0}+\int_{\sigma _{H}}\star \boldsymbol{\lambda }~,
\end{equation}
which constrains $\boldsymbol{\lambda }$ to satisfy
\begin{equation}
	\int_{\sigma _{H}}\star \boldsymbol{\lambda }=M_{0}\left( \frac{\partial g}{%
		\partial P}-\frac{V_{0}}{T_{0}}\frac{\partial g}{\partial S}\right) ~.
	\label{fixed lambda}
\end{equation}

To determine $\boldsymbol{\lambda}$ explicitly, we consider that the original description has a $M_{0}$ which is a generalized Komar energy.%
\footnote{The results of Section~\ref{sec: KadS}, combined with the development in Appendix~\ref{app:pot}, show that this original description exists.}%
In coordinate form, these quantities are expressed as
\begin{equation}
	M_{0}=-\frac{1}{8\pi }\int_{\sigma }(\nabla ^{\mu }\xi _{0}^{\nu }+\Lambda
	\omega_0 ^{\mu \nu })\mathrm{d}\sigma _{\mu \nu }~,\quad V_{0}=-\frac{1}{2}\int_{\sigma
		_{H}}\omega _{0}^{\mu \nu }\mathrm{d}\sigma _{\mu \nu }~,\quad \nabla _{\mu }\omega
	_{0}^{\mu \nu }=\xi _{0}^{\nu }~,
\end{equation}
where $\mathrm{d}\sigma_{\mu\nu}$ is the oriented surface element, and the following conservation equation for Killing vectors holds in vacuum:
\begin{equation}
	\nabla _{\mu }(\nabla ^{\mu }\xi _{0}^{\nu }+\Lambda \omega _{0}^{\mu \nu
	})=0~,\quad d\sigma _{\mu \nu }=\frac{\sqrt{-g}}{2}\tilde{\epsilon}_{\mu \nu
		\alpha \beta }\mathrm{d}x^{\alpha }\wedge \mathrm{d}x^{\beta }~.  \label{div = 0}
\end{equation}
In Equation~\eqref{div = 0}, $\tilde{\epsilon}_{\mu \nu \alpha \beta \,}$ represents the Levi--Civita symbol. Therefore, from Equation~\eqref{fixed
	lambda}, we fix $\boldsymbol{\lambda}$ with the coordinate expression
\begin{align}
\begin{split}
	&\lambda ^{\mu \nu }=\frac{1}{4\pi }\left( \frac{\partial g}{\partial P}-\frac{V_{0}}{T_{0}}\frac{\partial g}{\partial S}\right) (\nabla ^{\mu }\xi
	_{0}^{\nu }+\Lambda \omega _{0}^{\mu \nu })~,\quad \boldsymbol{\lambda }\equiv - \frac{1}{2}\lambda _{\mu \nu }\mathrm{d}x^{\mu }\wedge \mathrm{d}x^{\nu }~,
	\\
	&\star \boldsymbol{		\lambda }=-\frac{1}{2}\lambda _{\mu \nu }\mathrm{d}\sigma ^{\mu \nu }~.
\end{split}
        \label{eq:coordinate-expressions}
\end{align}
Furthermore, 
from Equation~\eqref{div = 0} (which implies $\star\mathrm{d}\star\boldsymbol{\lambda} = 0$), we derive the closure condition for~$\star\boldsymbol{\lambda}:\mathrm{d}{\star \boldsymbol{\lambda }=0}$.

We summarize the central result of this section. We have established that exact isohomogeneous transformations admit a geometric realization within the extended Iyer--Wald formalism. This implementation involves two key components:
\begin{itemize}

	\item A transformation of the Killing field [Equation~\eqref{K0 to K1}]; 
	
	\item A gauge transformation of the potential [Equation~\eqref{w0 to w1}].
\end{itemize}
The parameters $a\,$ and $\,b$ and the closed form $\star \boldsymbol{\lambda}$ are given by Equations~\eqref{a}, \eqref{b}, and \eqref{fixed lambda}. This construction guarantees that we can replace the operator $\delta $ with $\mathrm{d}$ in Equation~\eqref{transformed IW first law}.

Furthermore, the additional gauge freedom in our construction allows the thermodynamic description to naturally incorporate a temperature proportional to the surface gravity [see Equation~\eqref{a}]. This requirement was relaxed in~\cite{nos}. In the next section, we demonstrate that this novel framework has advantages over the implementation of the Iyer--Wald formalism as described in~\cite{xiao2024extended, gao2024general}.

\section{The Gauge of the Usual Kerr-Anti de Sitter Thermodynamics}
\label{sec: KadS}

As a key application, we demonstrate how conventional Kerr-anti de Sitter thermodynamics emerges as a special case of our generalized approach. The four-dimensional KadS spacetime is characterized by a mass parameter $m$, a rotation parameter $a$, and a negative cosmological constant $\Lambda$, describing an asymptotically anti-de Sitter rotating black hole. The line element, written in terms of the Boyer--Lindquist coordinates $(t,r,\theta,\phi)$, is
\begin{align}
\begin{split}
	\mathrm{d}s^{2}=-\frac{\Delta _{r}}{\rho ^{2}}\left( \mathrm{d}t-\frac{a\sin
		^{2}\theta \,\mathrm{d}\phi }{\Xi }\right) ^{2}+\frac{\Delta _{\theta }\sin
		^{2}\theta }{\rho ^{2}}&\left[ a\,\mathrm{d}t-\frac{\left( r^{2}+a^{2}\right)
		\,\mathrm{d}\phi }{\Xi }\right] ^{2}
	\\
	&+\frac{\rho ^{2}}{\Delta _{r}}\,\mathrm{d%
	}r^{2}+\frac{\rho ^{2}}{\Delta _{\theta }}\,\mathrm{d}\theta ^{2}\,,
\end{split}
	\label{Boyer-Lindquist Kerr-anti de Sitter}
\end{align}%
where 
\begin{equation}
\begin{split}
	&\Delta _{r} \equiv \frac{L^{2}+r^{2}}{L^{2}}\left( r^{2}+a^{2}\right)
	-2mr\,, \quad \Delta _{\theta }\equiv 1-\frac{a^{2}}{L^{2}}\cos ^{2}\theta \ , 
	\\
	&\rho ^{2} \equiv r^{2}+a^{2}\cos ^{2}\theta \,, \quad \Xi \equiv 1-\frac{a^{2}}{%
		L^{2}}\,, \quad L^{2}\equiv -\frac{3}{\Lambda }\,.
\end{split}
\end{equation}
The Lorentzian signature of the metric requires $\Xi >0$ for the validity of the $(t,r,\theta ,\phi )$ coordinate chart.

In a thermodynamic framework, it is useful to replace the set of
parameters $\{(m,a,\Lambda )\}$ by the set of thermodynamic variables $%
\{(S,J,P)\}$, 
\begin{equation}
	S=\frac{A}{4}=\frac{\pi \left( r_{+}^{2}+a^{2}\right) }{\Xi }~,\quad J=\frac{am}{%
		\Xi ^{2}}~,\quad P=\frac{3}{8\pi L^{2}}~,\quad m=\frac{\left( r_{+}^{2}+a^{2}\right)
		\left( L^{2}+r_{+}^{2}\right) }{2r_{+}L^{2}}~,  \label{h-ind}
\end{equation}%
with $r_{+}$ denoting the largest positive zero of the function $\Delta_{r}$
(the position of the \mbox{outer horizon).}

An initial thermodynamic description of KadS black holes was proposed by
Hawking~\cite{hawking1999rotation}. This development has a clear
geometric interpretation in terms of generalized Komar integrals
constructed from the Killing field $K$: 
\begin{equation}
	K=\partial _{t}+\Omega _{H}\partial _{\phi }~,  \label{K Hawking}
\end{equation}
where $\Omega_H$ is the scalar that ensures the null normalization of $K$ on the horizon.
For this choice, the extended Iyer--Wald formalism yields a variational
relation
\begin{equation}
\delta M_{H}=T_{H}\delta S+\Omega _{H}\delta J+V_{H}\delta P~,  \label{Hawking IW}
\end{equation}
with 
\begin{align}
\begin{split}
	&M_{H}\equiv\frac{m}{\Xi }~,\quad T_{H}\equiv\frac{\left( L^{2}+3r_{+}^{2}\right)
		r_{+}^{2}-a^{2}\left( L^{2}-r_{+}^{2}\right) }{4\pi L^{2}r_{+}\left(
		r_{+}^{2}+a^{2}\right) }~,
	\\
	&\Omega _{H}\equiv\frac{a\Xi }{a^{2}+r_{+}^{2}}~,\quad
	V_{H}\equiv\frac{4\pi }{3}\frac{r_{+}(r_{+}^{2}+a^{2})}{\Xi }~.
\end{split}
	\label{Hawking quantities}
\end{align}
However, it was noticed in~\cite{gpp} that $\delta M_{H}$ is
nonintegrable.%
\footnote{The cosmological constant is not treated as a thermodynamic variable in~\cite{gpp}. However, even if it is included, the resulting expression is still not a proper first law: $\text{d}M_{H}\neq T\,\text{d}S+\Omega_{H}\,\text{d}J+V_{H}\,\text{d}P\,$.} 
Hence, Hawking's relation~\eqref{Hawking IW} does not correspond to a proper first law.

Let us apply our approach to establish the gauge transformation in a concrete scenario. Since exact isohomogeneous transformations require a well-defined thermodynamic representation to generate new descriptions, Equations~\eqref{K Hawking}–\eqref{Hawking quantities} cannot serve as our starting point. However, a consistent version of Hawking’s approach can be constructed using nonexact homogeneous transformations, derived via the procedure in Appendix~\ref{app: non}. In this framework, Equation~\eqref{K Hawking} is replaced by
\begin{equation}
K_{A} \equiv \frac{\partial _{t}}{\sqrt{\Xi }}+\frac{\Omega _{H}}{\sqrt{\Xi }}\partial _{\phi }~.  
\label{K ATT}
\end{equation}
The index $A$ refers to ``alternative thermodynamic theory (ATT)'' for KadS, following the nomenclature introduced in~\cite{nos}.
As a result of the extended Iyer--Wald formalism, the variational equation of Equation~\eqref{IW gener} is a proper first law,
\begin{equation}
	\mathrm{d}M_{A}=T_{A}\mathrm{d}S+\Omega _{A}\mathrm{d}J+V_{A}\mathrm{d}P~,  \label{FL ATT}
\end{equation}
where the ATT quantities are defined as 
\begin{equation}
M_{A} \equiv \frac{M_{H}}{\sqrt{\Xi }}~, ~~
T_{A} \equiv \frac{T_{H}}{\sqrt{\Xi }} ~,~~ 
\Omega_{A} \equiv \frac{\Omega _{H}}{\sqrt{\Xi }} ~, ~~ 
V_{A} \equiv \frac{V_{H}}{\sqrt{\Xi }} ~,
\label{ATT quantities}
\end{equation}
and the Killing potential is implicitly given by 
\begin{equation}
V_{A}=\int_{\sigma _{H}}\boldsymbol{\omega }_{A}~.  
\label{omega ATT}
\end{equation}
In Appendix~\ref{app:pot} a proof that this $\boldsymbol{\omega }_{A}$ exists is presented.

From the ATT, new thermodynamic representations can be obtained for KadS from the EITs, using the procedure given in the previous section. This results in Equation~\eqref{transformed quantities} for $M_0 = M_A$, along with an arbitrary intensive function $g = g(S,J,P)$. Among these infinite possibilities, a more usual (and fairly explored) thermodynamic description of KadS black holes~\cite{caldarelli2000thermodynamics} can be obtained using the following exact isohomogeneous transformation from the ATT: 
\begin{equation}
(g,h)=\left( \frac{1}{\sqrt{\Xi }},M_{A}\right) ~, \quad 
\Xi (J,S,P)  = \bigg( 1 + \frac{8\pi}{3}\frac{J^{2}P}{M^{2}_{A}} \bigg)^{-1}~.
\label{eq:Xi}    
\end{equation}
This is referred to as the ``usual thermodynamic theory (UTT)'' in~\cite{nos} and denoted here by the index $U$. With the development presented in the previous section, we see that this transformation has a geometric counterpart, which is implemented in the extended \mbox{Iyer--Wald formalism.}

Combining Equation~\eqref{K0 to K1} with Equations~\eqref{a}, \eqref{b}, \eqref{K ATT}, and \eqref{eq:Xi}, we derive the transformed Killing field:%
\footnote{Although $K_{U}$ coincides with $K$ of \eqref{K Hawking}, its different
expansion is relevant since it is the first term inside parentheses that contributes to $\delta M_{U}$ in the Iyer--Wald formalism. Moreover, this is the vector that considers the angular velocity of the black hole with respect to infinity. Is is also interesting to note that this vector is no longer unique if we allow the temperature not to be proportional to the surface gravity. In this case, $K_{U}\neq K$, which is explored in a gauge-independent energy setup in~\cite{nos}.}
\begin{equation}
	K_{A}\longrightarrow K_{U}\equiv \left( \partial _{t}-\frac{a}{L^{2}}%
	\partial _{\phi }\right) +\left( \Omega _{H}+\frac{a}{L^{2}}\right) \partial
	_{\phi }~.  \label{KU}
\end{equation}
We also derive a gauge transformation, from Equations~\eqref{w0 to w1}, \eqref{fixed
	lambda} and \eqref{eq:coordinate-expressions}, expressed as 
\begin{equation}
	\omega _{A}^{\mu \nu }\longrightarrow \omega _{U}^{\mu \nu }=\sqrt{\Xi }%
	\omega _{A}^{\mu \nu }+\lambda ^{\mu \nu }~,\quad \lambda ^{\mu \nu }=\frac{a^{2}%
	}{3\sqrt{\Xi }}\left( \nabla ^{\mu }\xi _{A}^{\nu }+\Lambda \omega _{A}^{\mu
		\nu }\right) ~,\quad \xi _{A}\equiv \frac{\partial _{t}}{\sqrt{\Xi }}~.
	\label{omegaU}
\end{equation}

Following the extended Iyer--Wald formalism, the Killing field and gauge
transformations generate the first law given by 
\begin{equation}
	\mathrm{d}M_{U}=T_{U}\mathrm{d}S+\Omega _{U}\mathrm{d}J+V_{U}\mathrm{d}P~,
\end{equation}
where the UTT quantities are
\begin{equation}
M_{U} \equiv \frac{M_{H}}{\Xi } ~, ~~ 
T_{U} = T_{H}~, ~~ 
\Omega_{U} \equiv \Omega _{H}+\frac{a}{L^{2}}~, ~~ 
V_{U} \equiv V_{H}+\frac{4\pi }{3}a^{2}M_{U}~. 
\label{UTT quantities}
\end{equation}

The construction of the gauge $\boldsymbol{\omega }_{U}$ of Equation~\eqref{omegaU} is the main result of this section. It demonstrates that our extended Iyer--Wald formalism reproduces the usual KadS thermodynamic theory for the Killing vector $K_{U}$ of Equation~\eqref{KU}, without the need of any prior knowledge of the resulting theory. The method depends only on the existence of an original description (in our case, the ATT). Within our framework, the method of extracting an exact variation from a non-integrable first law in~\cite{xiao2024extended,gao2024general} corresponds to an appropriate gauge fixing.

We stress that we focus here on recovering the UTT because it remains a widely studied description in the literature. Nevertheless, ETIs can be used to generate infinite equivalent thermodynamic representations. This can be achieved, for instance, by generalizing Equation~\eqref{eq:Xi} to $(g,h) = (\Xi^n,M_A)$, with $n\neq0$. Furthermore, our generalized formalism reveals that the difference between thermodynamic and geometric volumes is a gauge term. For the UTT, this is the extra term that distinguishes $V_U$ and $V_H$ in Equation~\eqref{UTT quantities}.

\section{Final Remarks}
\label{sec: finals}

In previous work~\cite{nos}, we showed that exact isohomogeneous transformations constitute a fundamental structure for generating equivalent thermodynamic descriptions. In the present article, we analyze the formal mathematical foundations of these mappings. In particular, we show that EITs are contactomorphisms that include the well-known potentials obtained via Legendre transformations as a subset. As a central part of our study, we also demonstrate how these mappings can be connected to the gauge freedom present in formulations of black-hole thermodynamics based on a potential volume. 

Black-hole thermodynamics fundamentally differs from classical equilibrium thermodynamics in its reliance on functional variations rather than exterior derivatives, a distinction apparent in their respective formulations of the first law. In this regard, while the Iyer--Wald formalism is a powerful tool for deriving first laws in black-hole thermodynamics, it typically leads to non-integrable and non-homogeneous results, which obscure their connection to conventional thermodynamic first laws. We address this conceptual gap by constructing a framework that produces integrable first laws using EITs and a gauge-dependent extension of the Iyer--Wald formalism. An advantage of this approach is that it generalizes the infinitesimal Noether--Wald charge while preserving its conservation, which does not hold in extensions based on the (geometric) vector volume.

Our results demonstrate that exact isohomogeneous transformations, within the extended Iyer--Wald formalism, provide a systematic framework for identifying compatible gauge choices and appropriate Killing vector normalizations that yield fully integrable first laws. Equivalently, this can be interpreted as the statement that the geometric counterpart of EITs in the Iyer--Wald formalism corresponds to transformations in both the gauge and the horizon generator. The EIT-based extension, thus, establishes a robust connection between the geometric and thermodynamic descriptions of black holes.

A link for the different thermodynamic descriptions for Kerr-anti de Sitter black holes emerges from our analysis. As a concrete demonstration, we recovered standard KadS thermodynamics from an alternative description through a specific gauge selection. In our formalism, the procedure of extracting an exact differential of the black-hole mass from a non-integrable result is replaced by an appropriate gauge fixing. This, in turn, provides a new explanation for the well-known difference between the geometric and \mbox{thermodynamic volumes.}

In summary, our proposal, based on exact isohomogeneous transformations and a gauge-extended Iyer--Wald formalism, addresses longstanding integrability challenges in formulating a proper first law, and it offers a new perspective on KadS thermodynamics.
This work suggests several research directions. These include derivations of new KadS thermodynamic descriptions, and applications of EITs and the extended Iyer--Wald formalism to other gravitational theories containing free parameters.
A deeper understanding of the physical meaning behind the freedom to fix the gauge and select the Killing field also remains an important open question.
For instance, in order to connect this framework to frames of reference, it is necessary to supplement the conventional explanation that links the normalization of the Killing field to the Tolman factor for the black-hole temperature. Furthermore, the appropriate choice of gauge fixing may require a revised interpretation of KadS thermodynamics.
Investigations along these lines are currently underway.


\appendix

\section{Overview of Contact Manifold Description}
\label{app:contact}

The equilibrium states of a thermodynamic system can be identified
as Legendre submanifolds within a contact manifold, which is the
odd-dimensional analogue of a symplectic manifold. Given a 1-form $\alpha $ on a $(2n+1)$-dimensional manifold $\mathcal{T}$,
one can differentiably associate each point $p\in \mathcal{T}$ with a
hyperplane $\ker \alpha _{p}\subset T_{p}\mathcal{T}$, with $\dim (\ker
\alpha _{p})=2n$. This hyperplane $\ker \alpha _{p}$ is called a contact
plane, and the set of such hyperplanes defines a distribution $D\subset T \mathcal{T}$. Since, for any nonzero function $g:\mathcal{T}\rightarrow 
\mathbb{R}$, we have $\ker \alpha _{p}=\ker (g\alpha )_{p}$, we say that
1-forms differing by a conformal factor belong to the same equivalence
class. A map $f:\mathcal{T}\rightarrow \mathcal{T}$ such that the pullback
satisfies 
\begin{equation}
f^{\ast }\alpha =g\alpha  \label{classe}
\end{equation}
is called a contact transformation or contactomorphism. The special case $g=1$, for which the Legendre transformation is an example, is called a
strict contactomorphism.

If we further impose the condition
\begin{equation}
\alpha \wedge \omega ^{n}\neq 0~,\quad \omega =\mathrm{d}\alpha ~,
\label{alphan}
\end{equation}
the 2-form $\omega $ is non-degenerate on $D$, i.e., $\omega |_{D}\neq
0 $. Thus, at each point $p\in \mathcal{T}$, the tangent space decomposes as a direct sum $T_{p}\mathcal{T}=\ker \alpha _{p}\oplus \ker \omega _{p}$. The pair $(\mathcal{T},\alpha )$ is called a contact manifold, and $\alpha $ is the contact form. Darboux's theorem ensures that every contact manifold admits local coordinates, called canonical coordinates, in which 
\begin{equation}
	\alpha =\mathrm{d}z-\sum_{i=1}^{n}Y_{1}^{i}\mathrm{d}X_{i}~.  \label{stf}
\end{equation}

Regarding dynamics, condition~\eqref{alphan} implies $\dim (\ker \omega
_{p})=1$. That is, the contact structure $(\mathcal{T},\alpha )$ admits a
unique vector field $\xi$ satisfying
\begin{equation}
\imath_{\xi }\omega =0,\quad \alpha (\xi )=1~,  \label{reeb}
\end{equation}
where $\imath_{\xi }$ is the interior product. This vector field $\xi $ defines a decomposition \mbox{$T\mathcal{T}=L_{\xi }\oplus D$}, where $L_{\xi }$ is the
vertical subspace generated by $\xi $. It is important to note that this
decomposition is not unique, as it depends on the representative of the
class (\ref{classe}). The vector field $\xi $ is called the Reeb vector
field, and it plays the role of the Hamiltonian vector field, since its flow preserves $\omega $ (i.e., $\mathcal{L}_{\xi }\omega =0$). In canonical coordinates~\eqref{stf}, we have $\xi =\partial _{z}$. The equations of motion are invariant under transformations that preserve the 2-form $\omega\,$.

In the context of a thermodynamic description~\cite{bln}, we set $z=M_{0}$
as a thermodynamic potential, with $X_{i}$ representing the $n$ degrees of
freedom of the system. The set $(M_{0},X_{i},Y_{0}^{i})$ forms the canonical coordinates of a $(2n+1)$-dimensional manifold $\mathcal{T}$, referred to as the thermodynamic phase space. In this case,
\begin{equation}
\alpha =\mathrm{d}M_{0}-\sum_{i=1}^{n}Y_{0}^{i}\mathrm{d}X_{i}~,  \label{c1}
\end{equation}
where $X_{1}=S$ and $Y_{0}^{1}=T$. The kernel of $\alpha $
represents the equilibrium states of the system. Since $\alpha $ and $\eta
=g\alpha $, with $g\neq 0$, have the same kernel, they describe the same
equilibrium states, that is, they define equivalent thermodynamic
descriptions. However, the freedom in the choice of canonical coordinates
allows for different thermodynamic formulations associated with the same
equilibrium states.

\section{Nonexact Isohomogeneous Transformations}

\label{app: non}

The homogeneity of the black-hole mass can be recovered including the
functional variation of all free parameters, such as the cosmological
constant. It follows that a Smarr formula can be established given a
particular choice of normalization for the Killing field that generates the
horizon. However, the direct application of the Iyer--Wald formalism to the
resulting variables may yield a first law that is not integrable, 
\begin{equation}
\delta M_{1}=T_{1}\delta S+\sum_{i=2}Y_{1}^{i}\delta X_{i}~,  \label{not FL}
\end{equation}
where $\delta M_{1}$ is not an exact form $\mathrm{d}M_{1}$, and the index $1$ is used for convenience (to be explained shortly). In some interesting
cases, we can find an intensive function $g$, which will be treated as an
independent variable (for the moment), so that Equation~\eqref{not FL} can be
rewritten in the differential-form version 
\begin{equation}
T_{1}\mathrm{d}S+\sum_{i=2}Y_{1}^{i}\mathrm{d}X_{i}=\mathrm{\ d}M_{1}-\frac{
		M_{1}}{g}\mathrm{d}g~.  \label{h=0}
\end{equation}
This is the result of Equation~\eqref{int} for a nonexact isohomogeneous
transformation with $h=0$, which can be simplified to 
\begin{equation}
\mathrm{d}\frac{M_{1}}{g}=\frac{T_{1}}{g}\,\mathrm{d}S+\sum_{i=2}\frac{Y_{1}^{i}}{g}\,\mathrm{d}X_{i}~,  \label{FL hzero}
\end{equation}
and we say that this theory is the thermodynamic version of Equation~\eqref{not
	FL}.

Thus, the homogeneous function $M_{1}$ is not a thermodynamic potential in
the variables $T_{1},\left\{ Y_{1}^{i}\right\} $. However, it is
straightforward to define a legitimate thermodynamic \mbox{potential with}
\begin{equation}
	M_{0}=\frac{M_{1}}{g}~,\ T_{0}=\frac{T_{1}}{g}~,\ Y_{0}^{i}=\frac{Y_{1}^{i}}{%
		g}~.  \label{hzero}
\end{equation}%
Geometrically, the implementation of~\eqref{hzero} can be seen as a
renormalization of the Killing field that generates the thermodynamics. If
Equation~\eqref{not FL} is obtained from the Killing vector $K_1$ and the
thermodynamic description of Equation~\eqref{FL hzero} from $K_0$, then the
geometric implementation of Equation~\eqref{hzero} can be seen as the
transformation 
\begin{equation}
	K_{0}\longrightarrow g\,K_{0}=K_{1}~.
\end{equation}

As an important example, we cite Hawking's work~\cite{hawking1999rotation}, which defines thermodynamic quantities that, from the extended Iyer--Wald formalism, result in Equation~\eqref{Hawking IW}, with a Smarr formula 
\begin{equation}
	M_H = 2 T_H S + 2 \Omega_H J - 2 V_H P~,  \label{kaH}
\end{equation}
which can be identified with the second relation of Equation~\eqref{int}. This is
a case where the Smarr formula does not correspond to an Euler relation, as
commented in Section~\ref{sec it}.

As shown in~\cite{nos}, this variational relation can be written in the form~\eqref{h=0} as 
\begin{equation}
	\mathrm{d}M_H - \frac{M_H}{\sqrt{\Xi}} \mathrm{d} \sqrt{\Xi} = T \mathrm{d}S
	+ \Omega_H \mathrm{d}J + V_H \mathrm{d} P~,
\end{equation}
with the thermodynamic quantities defined in Section~\ref{sec: KadS}. From
our considerations, there is a proper thermodynamic description given by
Equations~\eqref{FL hzero} and~\eqref{hzero} with $g = \sqrt{\Xi}$. This is the
alternative thermodynamic theory of KadS of Equation~\eqref{FL ATT}, with
renormalized Killing field~\eqref{K ATT}. Thus, the ATT is the thermodynamic
version of Hawking's approach.

We conclude that nonexact isohomogeneous transformations are a powerful tool for establishing proper thermodynamic versions of non-integrable first laws that can be written in the form~\eqref{h=0}. Moreover, their usefulness is not limited to these cases of $h=0$. Broader applications can be found in~\cite{nos}, where, for instance, it is shown how the usual KadS thermodynamics can be obtained from Hawking's approach.

\section{Matching Potential and Vector Volumes}

\label{app:pot}

It is known that $V_{A}$ in \eqref{omega ATT} can be written as a vector
volume of the region between the singularity and the horizon~\cite{nos}. We
want to show that there exists a gauge such that $V_{A}$ is also a potential
volume. In greater generality, we want to find the gauge such that 
\begin{equation}
	\lim_{\sigma \rightarrow 0}\int_{\sigma }^{\sigma _{H}}\imath_{v}(\star
	1)=\int_{\sigma _{H}}\star \boldsymbol{\omega }~,\quad \imath_{v}(\star 1)=\mathrm{d}%
	\star \boldsymbol{\omega }~.
\end{equation}%
For that, take an auxiliary closed (but not exact) differential form $%
\boldsymbol{\theta }$ such that $\theta \neq 0$, and~define 
\begin{equation}
	\boldsymbol{\theta }^{\prime }\equiv -\frac{\omega (\sigma \rightarrow 0)}{%
		\theta }\boldsymbol{\theta }~,\quad \omega (\sigma \rightarrow 0)\equiv
	\lim_{\sigma \rightarrow 0}\int_{\sigma }\star \boldsymbol{\omega }~,\quad
	\theta \equiv \int_{\sigma }\boldsymbol{\theta }~.
\end{equation}%
Then, because there exists a $\boldsymbol{\theta }$ in KadS, 
\begin{equation}
	V_{v}=\lim_{\sigma \rightarrow 0}\int_{\sigma }^{\sigma _{H}}\imath_{v}(\star
	1)=\lim_{\sigma \rightarrow 0}\int_{\sigma }(\star \boldsymbol{\omega }+%
	\boldsymbol{\theta }^{\prime })+\int_{\sigma _{H}}(\star \boldsymbol{\omega }%
	+\boldsymbol{\theta }^{\prime })=\int_{\sigma _{H}}(\star \boldsymbol{\omega 
	}+\boldsymbol{\theta }^{\prime })~,
\end{equation}
which completes the proof since $\boldsymbol{\theta }^{\prime }$ is closed.

\section*{Funding}

T.~L.~C. acknowledges the support of Coordena\c{c}\~ao de Aperfei\c{c}oamento de Pessoal de N\'{\i}vel Superior (CAPES) -- Brazil, Finance Code 001.
C.~M. is supported by Grant No.~2022/07534-0, S\~ao Paulo Research Foundation (FAPESP), Brazil.

\begin{acknowledgments}
	
M. C. B. thanks Professor Orfeu Bertolami at University of Porto for hosting part of this study.
	
\end{acknowledgments}


\begin{thebibliography}{99}
	
\bibitem{bln} Bravetti, A.; Lopez-Monsalvo, C.S.; Nettel, F. Contact
Symmetries and Hamiltonian Thermodynamics. \textit{Ann. Phys.} \textbf{2015}, \textit{361},
377--400.

\bibitem{iw} Iyer,~V.; Wald, R. Some Properties of Noether
Charge and a Proposal for Dynamical Black Hole Entropy. \textit{Phys. Rev. D} \textbf{1994}, \textit{50}, 846--864.

\bibitem{wald1993black} Wald, R. Black hole entropy is the Noether charge. \textit{Phys. Rev. D} \textbf{1993}, \textit{48}, R3427. 

\bibitem{hs} Hajian, K.; Sheikh-Jabbari, M.M. Solution
phase space and conserved charges: A general formulation for charges
associated with exact symmetries. \textit{Phys. Rev. D} \textbf{2016} \textit{93}, 044074.
arXiv:1512.05584

\bibitem{xiao2024extended} Xiao, Y.; Tian, Y.; Liu, Y.X. Extended
black hole thermodynamics from extended Iyer--Wald formalism. \textit{Phys. Rev. Lett.} \textbf{2024}, \mbox{\textit{132}, 021401}. 
arXiv:2308.12630

\bibitem{kastor2009enthalpy} Kastor, D.; Ray, S.; Traschen, J. Enthalpy and the mechanics of adS black holes. \textit{Class. Quant. Grav.} \textbf{2019}, \textit{26}, 195011. 
    arXiv:0904.2765

\bibitem{kmt} Kubiznak, D.; Mann, R.B.; Teo, M. Black hole chemistry:
thermodynamics with Lambda \textit{Class. Quantum Grav.} \textbf{2017},  \textit{34}, 063001.
arXiv:1608.06147

\bibitem{nos} Campos, T.; Baldiotti, M.C.; Molina, C. Generating
Kerr-anti-de Sitter thermodynamics.\textit{ Phys. Rev. D} \textbf{2024}, \textit{110}, 024049. 
arXiv:2407.09610

\bibitem{gao2024general} Gao, Y.; Di, Z.; Gao, S. General mass
formulas for charged Kerr--adS black holes. \textit{Phys. Scr.} \textbf{2024}, \textit{99}, 095022. 
arXiv:2304.10290

\bibitem{gpp} Gibbons, G.W.; Perry, M.J.; Pope, C.N. The First Law of Thermodynamics for Kerr-Anti-de Sitter Black Holes. \textit{Class. Quant. Grav.} \textbf{2005}, 
\textit{22}, 1503. 
arXiv:hep-th/0408217

\bibitem{hawking1999rotation} Hawking, S.W.; Hunter, C.J.; Taylor-Robinson, M.M. Rotation and the AdS/CFT correspondence. \textit{Phys.
	Rev. D} \textbf{1999}, \textit{59}, 064005. 
    arXiv:hep-th/9811056

\bibitem{bfm} Baldiotti, M.C.; Fresneda, R.; Molina. C. A Hamiltonian
approach to Thermodynamics. \textit{Ann. Phys.} \textbf{2016}, \textit{373}, 245--256. 
arXiv:1604.03117

\bibitem{brav} Bravetti, A. Contact geometry and thermodynamics. \textit{Int.
	J. Geom. Meth. Mod. Phys.} \textbf{2018}, \textit{16}, 1940003.

\bibitem{bfm2} Baldiotti, M.C.; Fresneda, R.; Molina, C. A Hamiltonian
approach for the Thermodynamics of AdS black holes. \textit{Ann. Phys. } \textbf{2017}, \textit{382}, 22--35. 
arXiv:1701.01119

\bibitem{fbfm} Fontana, W.B.; Baldiotti, M.C.; Fresneda, R.;  Molina, C. Extended quasilocal Thermodynamics of Schwarzchild-anti de Sitter black holes. \textit{Ann. Phys.} \textbf{2019}, \textit{411}, 167954. 
arXiv:1806.05699

\bibitem{gibbons1977action} Gibbons, G.W.; Hawking, S.W. Action
integrals and partition functions in quantum gravity. \textit{Phys. Rev. D} \textbf{1977}, \textit{15}, 2752.

\bibitem{Wei} Weinhold, F. Metric geometry of equilibrium
thermodynamics. \textit{J. Chem. Phys.} \textbf{1975}, \textit{63}, 2479.

\bibitem{cfn} Couch, J.; Fischler, W.; Nguyen, P.H. Noether charge, black hole volume, and complexity. \textit{J. High Energy Phys.}  
\textbf{2017}, \textit{2017}, 119. 
arXiv:1610.02038

\bibitem{bl} Ballik, W.; Lake, K. The Vector Volume and Black Hole. \textit{Phys. Rev. D} \textbf{2013}, \textit{88}, 104038. 
arXiv:1310.1935

\bibitem{bardeen1973four} Bardeen, J.M.; Carter, B.; Hawking, S.W. The four laws of black hole mechanics. \textit{Commun. Math. Phys.} \textbf{1973}, \textit{31}, 161.

\bibitem{dolan2011pressure} Dolan, B.P. Pressure and volume in the
first law of black hole thermodynamics. \textit{Class. Quant. Grav}. \textbf{2011}, \textit{28},
235017. 
arXiv:1106.6260

\bibitem{dolan2012pdv} Dolan, B.P. Where is the PdV in the first law of black hole thermodynamics? In \textit{Open Questions in Cosmology}, 1st ed.; Olmo, G.J., Ed.; IntechOpen: Rijeka, Croatia, 2012; pp. 291--315. 
 arXiv:1209.1272

\bibitem{cvetivc2011black} Cveti{\v{c}}, M.; Gibbons, G.W.;  Kubiz{\v{n}}{\'{a}}k, D.; Pope, C.N. Black hole enthalpy and an entropy inequality
for the thermodynamic volume. \textit{Phys. Rev. D} \textbf{2011}, \textit{84}, 024037.
arXiv:1012.2888

\bibitem{rodriguez2022first} Rodr{\'\i}guez, N.H.; Rodriguez, M.J. First law for Kerr Taub-NUT AdS black holes. \textit{J. High Energy Phys.} \textbf{2022}, \textit{2022}, 44. 
arXiv:2112.00780

\bibitem{caldarelli2000thermodynamics} Caldarelli, M.; Cognola, G.; Klemm, D. Thermodynamics of Kerr-Newmann-AdS black holes and conformal
field theories. \textit{Class. Quant. Grav.} \textbf{2000}, \textit{17}, 399.
arXiv:hep-th/9908022

\end{thebibliography}
\end{document}